Peculiarities of the defect model for the mixed mobile ion effect in mixed cation glasses

Vladimir Belostotsky

1005 Curtis Place, Rockville, MD 20852

vladbel@erols.com

ABSTRACT

Specific features of mixed cation effect in ternary glasses whose composition involves alkali and non-alkali unicharged ions and/or alkaline earth cations are explored within the context of the defect model for the mixed mobile ion effect (V. Belostotsky, J. Non-Cryst. Solids (2007, 2010)). In the mixed alkaline earth glasses mixed cation effect is observed as in ionic diffusivity as well as in the glass transition temperature reflecting the fact that movement of dissimilar ions of unequal size is accompanied by the formation of intrinsic defects in glass forming matrix when mobile cations enter foreign sites. It is qualitatively similar to that in mixed alkali glasses although quantitatively it is much less pronounced due to lower mobility of the alkaline earth cations. In mixed cation glasses whose composition involves dissimilar mobile ions with ionic radii close to each other the 'classical' mixed mobile ion effect is not observed in a sense we face it in mixed alkali glasses. In alkali alkaline earth glasses like $xNa_2O\text{-}(1\text{-}x)CaO\text{-}SiO_2$ and $xK_2O\text{-}(1\text{-}x)BaO\text{-}SiO_2$ the energy landscape is comprised of the regular unicharged alkali sites and double-charged alkaline-earth sites, and mobile cations of both types can utilize either vacant sites without their rearrangement. This leads to increase in apparent diffusivity of alkaline-earth species when their concentration decreases which is opposite to what is reported for mixed alkali glasses. The mixed cation effect in the transition temperature of such glasses is not observed which is a clear indication that the site rearrangement involving generation of intrinsic defects does not occur when a mobile ion enters a foreign site. Similar picture is observed in silver-sodium phosphate glasses where mixed cation effect in ionic conductivity is anomalously low because the dissimilar unicharged ions, $Na^+$ and $Ag^+$, have almost equal ionic radii and can utilize either vacant ionic sites without their rearrangement and intrinsic defect formations in the nearest coordination environment.

1. INTRODUCTION

The defect model for the mixed mobile ion effect (MMIE) [1,2] has demonstrated that the non-linear compositional variations of various physical properties of mixed alkali glasses and other mixed ionic conductors arise from intrinsic structural defects generated in network-forming matrix in response to the simultaneous migration of two or more dissimilar mobile ions of unequal size. An approach introduced in [1,2] for mixed alkali glasses is generally applicable to any mixed cation and anion systems. However, in spite of obvious similarities, MMIE in mixed alkaline earth, alkali alkaline earth and other mixed cation glasses (which will be further referred to as 'mixed cation effect', MCE) has some specific features. Ionic conductivity in alkali alkaline earth glasses is driven by the alkali content, whereas mixed alkaline earth glasses in the absence of alkali impurities are not ionic conductors [3] and the MCE in this case has to be described in terms of ionic diffusivities. On the other hand, in the presence of unicharged alkali ions in a number of alkali alkaline earth systems double-charged alkaline earth cations demonstrate *greater* diffusivities than in binary alkaline earth glasses [4,5]. In addition, in such glasses MCE in the transition temperature is not observed [6,7]. Recent studies of mixed cation phosphate glasses of general composition *$xAg_2O$-$(1-x)Na_2O$-$P_2O_5$* where the dissimilar unicharged ions, *$Ag^+$* and *$Na^+$*, have almost equal size demonstrate anomalously low mixed cation effect in ionic conductivity [8]. All these otherwise deviations from the expected behavior have not received an adequate explanation in the literature. Thus, the main objective of the present work is to discuss and account for the peculiarities of the MCE in various mixed cation glasses in light of the approach introduced in [1,2] which is given in the next section.

Experimental data to be discussed here have been compiled from the literature. Unfortunately, mixed alkaline, alkali alkaline and other mixed cation glasses have attracted much less attention of the researches than mixed alkali glasses, and the literature data on these systems are relatively scarce. There are only few papers where the properties of such glassy systems are investigated systematically [3-6,9-14]. Much of the data discussed here are obtained from the cation diffusivity studies. Why mechanical loss measurements are not taken into consideration is also discussed in the section that follows.

In this way, the present work can be viewed as an important addendum to the defect model for the mixed mobile ion effect.

## 2. MIXED MOBILE ION EFFECT IN MIXED CATION GLASSES

In mixed alkaline earth glasses, the diffusivity of the dissimilar cations is characterized by activation energies greater than those in the end-point single alkaline earth glasses [10], and the glass transition temperature of such glasses demonstrates a deviation from linearity with lower $T_g$ at intermediate alkaline earth compositions than it is expected from the linear interpolation of the transition temperatures of corresponding single alkaline earth glasses [11-13]. Thus, with an exception of ionic conductivity, the MCE in mixed alkaline earth glasses is qualitatively similar to that in mixed alkali glasses although quantitatively it is less pronounced [10]. This is apparently due to much lower mobility of the double charged alkaline earth ions: fewer alkaline earth cations can move into foreign sites and cause their rearrangement involving the generation of intrinsic defects in the adjacent glass matrix.

The MCE on ionic conductivity of alkali alkaline earth glasses is difficult to elicit because only alkali cations contribute to ionic conductivity whereas alkaline earth species remain virtually immobile within the time window of hopping process of the alkali ions [10]. Since the ionic conductivity is concentration-dependent entity, it is difficult to determine which part of the conductivity reduction shall be referred to the alkali content decrease when alkali oxide is gradually replaced by alkaline earth oxide at fixed concentration of the glass former and which one is due to trapping of a part of alkali ions in rearranged former alkaline earth sites acting as high energy anion traps.

To establish whether there is MCE in alkali alkaline earth glass, $T_g$ concentration dependence has to be taken into consideration. Its departure from linearity at intermediate alkali alkaline earth compositions can serve as an indicator of the weakening the glass network due to formation of intrinsic defects on ion mixing. Fig. 1 shows a remarkable MCE in $T_g$ of two alkali alkaline earth glasses where dissimilar cations of unequal size are mixed given in comparison with MCE in $T_g$ of mixed alkali glass.

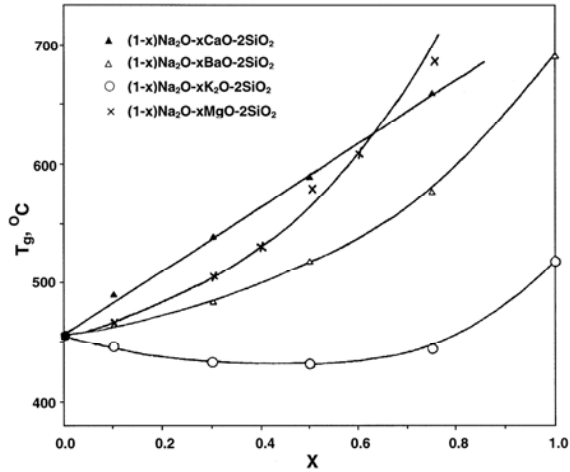

Fig. 1. Comparison of the effect of $Na^+$, $Ca^{2+}$, $Ba^{2+}$, $K^+$, and $Mg^{2+}$ on the glass transition temperature, $T_g$, of mixed *Na-Ca, Na-Ba, Na-K,* and *Na-Mg* disilicate glasses. (Compiled from Ref. 6).

Alkali alkaline earth glasses whose composition involves dissimilar cations with ionic radii close to each other, do not exhibit a 'classical' mixed mobile ion effect in a sense we face it in mixed alkali glasses. The typical examples are the glasses of compositions *(1-x)Na$_2$O-xCaO-SiO$_2$* and *(1-x)K$_2$O-xBaO-SiO$_2$* whose alkali and alkaline earth ionic radii are summarized in the Table 1. In such glassy systems, the energy landscape is comprised of the regular uncharged alkali sites and double-charged alkaline-earth sites, and mobile cation of any type can utilize either vacant site without its immediate rearrangement and intrinsic defect generation in adjacent network-forming matrix. It could be suggested here that careful self-diffusion measurements for the alkali and alkaline earth species, as is done by Frischat for mixed alkali glasses [15], would reveal two distinct subsets of sites for each type of cations with very different activation energies: alkali and alkaline earth cations residing in their regular uncharged alkali sites and regular double-charged alkaline earth sites, respectively, uncharged alkali cations trapped in double-charged sites vacated by alkaline earth cations, and double-charged alkaline earth cations occupying uncharged former alkali sites. Former alkaline earth sites act as high-energy anion traps for alkali cations increasing in this way apparent activation energy of the overall alkali content whereas Coulomb energy binding a fraction of double-charged alkaline-earth cations to former alkali sites is roughly half of that of the regular alkaline-earth sites, which reduce apparent activation energy of the overall alkaline-earth content. This is observed as an enhanced diffusivity of alkaline-earth species when they are substituted by alkalis, which is contrary to what is found in mixed alkali glasses [3-5,14,16]. At the same time, the MCE in $T_g$ of such glasses is not detected [6,7]. As can be seen from the Fig.1, the transition temperature of *(1-x)Na$_2$O-xCaO-2SiO$_2$* glass demonstrates a linear concentration dependence [6] which is a clear indication that the site rearrangement and generation of intrinsic defects does not occur.

TABLE I. Most frequent cation coordination numbers (*CN*), ionic radii ($r_i$), and cation – oxygen internuclear distances ($r_{i\text{-}O}$) in silicates [17-19].

| Ion | CN | $r_i$, pm | $r_{i\text{-}O}$, pm | Ion | CN | $r_i$, pm | $r_{i\text{-}O}$, pm |
|---|---|---|---|---|---|---|---|
| *Na* | 5 | 100 | 237 | *Ca* | 6 | 100 | 237 |
|  | 6 | 102 | 239 |  | 7 | 106 | 243 |
| *K* | 6 | 138 | 275 | *Ba* | 6 | 134 | 271 |
|  | 7 | 146 | 283 |  | 7 | 138 | 275 |

Mixing of $Ag^+$ and $Na^+$ uncharged cations in phosphate glasses also demonstrate dramatic departure from the expected behavior. Hall et al. have discovered that the MCE on ionic conductivity of $xNa_2O\text{-}(1\text{-}x)Ag_2O\text{-}P_2O_5$ glasses is anomalously low [8]. Studies of the structure of these glasses show that dissimilar cations, $Na^+$ and $Ag^+$, have almost equal ionic radii and occupy structurally similar sites. Mobile cations of both types can freely move into the foreign sites without their rearrangement and formation of intrinsic defects in adjacent network forming matrix. Systematic data on the glass transition temperature for these glasses is not available so far, however it can be predicted that MCE in $T_g$ of these glasses is also small.

3. WHY MECHANICAL LOSS SPECTROSCOPY IS NOT A SUITABLE TECHNIQUE FOR OBTAINING DATA ON THE MOBILITY OF ALKALINE EARTH CATIONS IN MIXED ALKALINE EARTH AND ALKALI ALKALINE EARTH GLASSES

The studies of mechanical losses provide valuable information about various relaxation processes in glasses. The position, magnitude and shape of the mechanical loss maxima, measured as a function of temperature or frequency, are defined by the type, concentration, and activation energy of structural units responding to the mechanical perturbation.

However, in contrast to, say, secondary ion mass spectroscopy, mechanical spectroscopy does not provide direct identification of the chemical species or structural moieties movement of which is relevant to the mechanical losses of glasses.

Supplemental investigations of electrical and dielectric relaxation, diffusion etc. are often employed to obtain a basis for comparison in order to identify actual mechanisms responding to the mechanical perturbation [20]. Frequently, close proximity in activation energy is the most conclusive argument for ascribing maxima in mechanical losses. As a sole reason for assignment, though, this approach is not well justified because various relaxation processes in glass may have close activation energies.

In order to verify the relevance of the mechanical loss maxima assignments it is not necessary, however, to rely solely on the external methods. In some instances, mechanical loss spectra themselves can be used for comparison if the origin of at least one of loss peaks is well established. It is based on the fact that the response of a material to a perturbation (the magnitude of the relaxation) is proportional to the concentration of structural units responding to a perturbation, and different relaxation processes giving rise to various loss peaks have essentially identical macroscopic mathematical description [21,22].

As an example of such analysis, we compare mechanical loss spectra of two glasses, binary sodium silicate glass *$Na_2O$-$3SiO_2$* and mixed alkali glass *$0.5Na_2O$-$0.5K_2O$-$3SiO_2$* having equal concentrations of glass former and alkali cations.

The studies of the mechanical losses of single alkali glasses reveal two sub-$T_g$ characteristic maxima in a fixed frequency variable temperature scans [23-27]. In silicate glasses, the first maximum typically appears below $0^o$C, and the second one is observed between $100^o$C and $300^o$C. There is strong consensus in the literature that a low temperature relaxation maximum is caused by the stress-induced movement of alkali ions. This attribution is based on the agreement between the activation energy calculated for this maximum and that for the dc ionic conductivity and alkali diffusion [25-28]. In the mechanical loss spectrum of *$Na_2O$-$3SiO_2$* glass displayed in Fig.2 low-temperature alkali peak maximum is observed at -33°C [21], and the activation energy, $E_a$, obtained for this peak is ~ 0.7 eV.

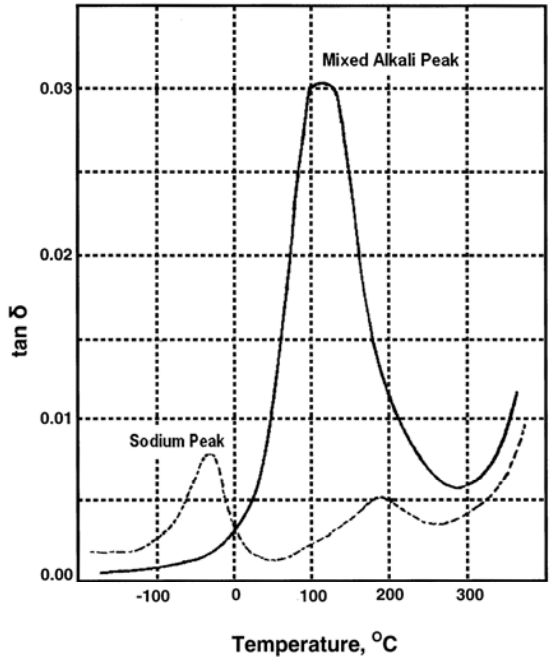

Fig. 2. Mechanical loss spectra of $Na_2O$-$3SiO_2$ (dashed line) and $0.5Na_2O$-$0.5K_2O$-$3SiO_2$ (solid line) glasses. (Compiled from Ref. 31).

The total number density of sodium ions in this system, $n_o$, is ~ $1.2 \times 10^{22}$ cm$^{-3}$. As is well known, only a fraction of alkali content, $n$, can overcome potential barrier $E_a$ and respond to the mechanical perturbation. This quantity can be calculated from the Arrhenius equation [22]:

$$n = n_O \exp\{-E_a / k_B T\} \qquad (1)$$

where $k_B$ is the Boltzmann constant and $T$ is temperature. Inserting the parameters of the low temperature maximum in Equation (1), we obtain the mobile fraction of sodium cations, $n = 2.3 \times 10^7$ cm$^{-3}$. Again, this is the very fraction of sodium ions responding to the perturbation and giving rise to the low temperature alkali peak in mechanical losses of *Na$_2$O-3SiO$_2$* glass, and the magnitude above background of this peak is proportional to *n*.

The structural origin of the second, high temperature maximum of mechanical losses of the single alkali glass has been ascribed to the movement of non-bridging oxygen (NBO) ions [27,29,30], although other opinions have been offered [32-38]. In the spectrum displayed in Fig.2 high-temperature peak maximum is observed at 182$^O$C, and the activation energy for this peak is approximately 1.4 eV [25].

The origin of the high-temperature peak can be verified by computing the mobile fraction of oxygens in this system and comparing it with the calculated mobile fraction of the sodium cations. The total number density of oxygens in this glass is $4.2 \times 10^{22}$ cm$^{-3}$. Then, the quantity in question calculated from Equation (1) is $1.3 \times 10^7$ cm$^{-3}$. As can be seen from Fig. 2, the magnitude above background of the high-temperature peak is ~ 40% lower than that of the sodium peak which perfectly matches to the ratio of the mobile fractions of the sodium and oxygen. This makes it clear that the high-temperature maximum in mechanical losses of single alkali glasses is indeed caused by the movement of oxygen-related structural moieties.

When one alkali modifier is gradually replaced by another alkali species, the low temperature alkali peak rapidly diminishes in magnitude and shifts to higher temperatures. The most profound effect of the alkali mixing on the mechanical loss spectra is the development of so-called 'mixed alkali peak'. This maximum in mechanical losses is typically observed between 100$^O$C and 300$^O$C. With increase in the concentration of the second alkali, it rapidly increases in magnitude and moves to lower temperatures. Typically, mixed alkali glasses with close concentrations of dissimilar alkali ions exhibit only this sub-$T_g$ maximum in mechanical losses due to its large magnitude. Our primary point of interest here is this 'mixed alkali peak' whose origin has been debated in the literature since the first report on mechanical losses (internal friction) of mixed alkali glasses was published. In 1969, Shelby and Day [31] have ascribed the 'mixed alkali peak' to the cooperative motion of unlike alkali ions, specifically to the reorientation of electrically neutral elastic dipoles. This viewpoint dominates the literature ever since even though it has not received general acceptance.

Mechanical loss spectrum of mixed alkali *0.5Na$_2$O-0.5K$_2$O-3SiO$_2$* glass is displayed in the Fig. 2. The comparison of the 'mixed alkali peak' with the alkali peak of *Na$_2$O-3SiO$_2$* glass shows that its magnitude above background is 5 times higher than that of the latter one, therefore the fraction of atomic particles, $n^{'}$, movement of which gives rise to this maximum should be $1.5 \times 10^8$ cm$^{-3}$. The 'mixed alkali peak' maximum is observed at 115$^O$C, and the activation energy, $E_a^{'}$, calculated for this maximum is ~ 1.1 eV [31]. Total number density of atomic particles giving rise to the 'mixed alkali peak', $n_O^{'}$, calculated using equation

$$n_O^{'} = n^{'} \exp\{E_a^{'} / k_B T\} \qquad (2)$$

is ~ $4.2 \times 10^{22}$ cm$^{-3}$.

As can be seen, 'mixed alkali peak' in mechanical losses of mixed alkali glass is caused by the mobile fraction of a glass constituent whose total number density is 3.5 times greater than that of the alkali one. The only glass constituent that meets this condition is oxygen. Thus, the data analysis performed here makes it evident that the 'mixed alkali peak' in mechanical losses is not related to a movement of alkali cations and is, in fact, the enlarged in magnitude maximum assigned to the movement of the oxygen-related structural moieties [1].

The quantitative data analysis outlined above for mixed alkali glasses can be employed to the assessment of the mechanical loss spectra of mixed alkaline earth and alkali alkaline earth glasses.

Fig. 3 shows the sub-$T_g$ spectrum of mechanical losses of $Na_2O$-$2CaO$-$4SiO_2$ glass obtained at a frequency 1 Hz [39]. Other alkali alkaline earth silicate glasses exhibit similar spectra of mechanical relaxation [39,40]. This spectrum is selected for the analysis because all its parameters are known. No doubt exists regarding the origin of the low temperature peak which linked to the movement of sodium ions. It is observed at $120^O C$, and activation energy obtained for this peak is 1.025 eV, which coincides with activation energy of dc conductivity of this glass. The second peak at 420°C with activation energy 1.88 eV the authors attributed to the movement of calcium cations. The number density of sodium ions in this glass is $3.55 \times 10^{21}$ cm$^{-3}$, the calcium amounts to $7.1 \times 10^{21}$ cm$^{-3}$, and the concentration of oxygen is $3.9 \times 10^{22}$ cm$^{-3}$.

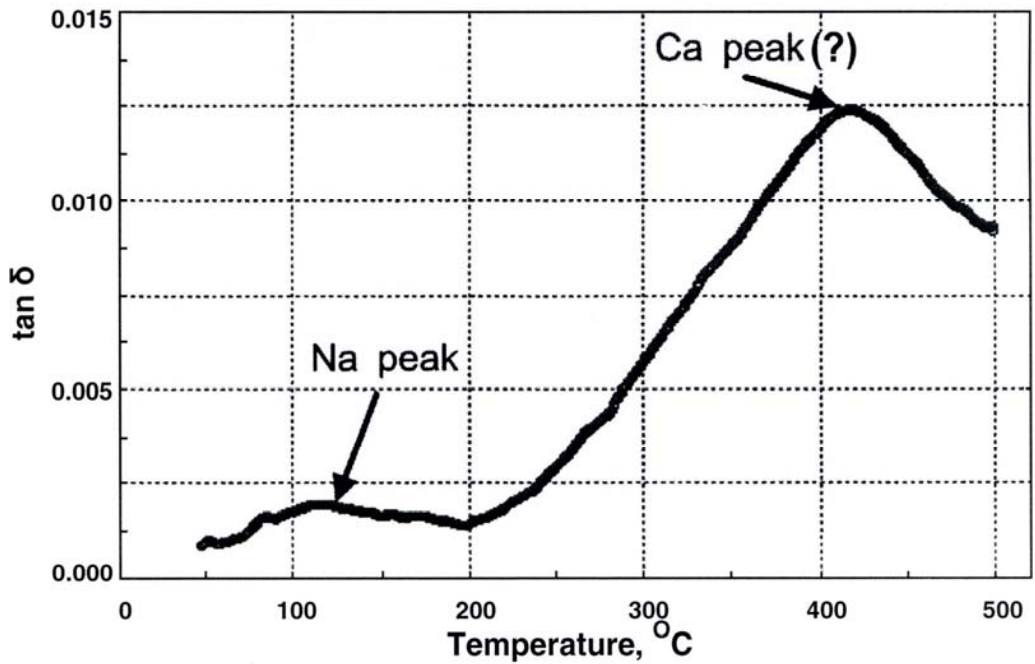

Fig. 3. The mechanical loss spectrum of $Na_2O$-$CaO$-$4SiO_2$ glass. (Redrawn from Ref. 39, p. 51).

Calculations using Equation (1) show that the mobile fraction of sodium ions giving rise to the low temperature peak reaches $2.45 \times 10^{8}$ cm$^{-3}$ whereas the mobile fraction of calcium at 420°C does not exceed $1.44 \times 10^{8}$ cm$^{-3}$. As can be seen, the low temperature sodium peak is poorly resolved in the spectrum, so even smaller mobile fraction of calcium cations which is almost two times lower than that of the sodium one is too minute for being observable in mechanical loss spectrum. This makes it immediately obvious that the movement of potassium cations can not cause the relaxation peak at 420°C. If we assume, however, that this maximum in mechanical losses is caused by the movement of the oxygen-related structural moieties, their mobile fraction calculated from the Equation (1) would be $\sim 1 \times 10^{9}$ cm$^{-3}$ or 4 times larger than that of the sodium one. This quantity is consistent with the observed spectrum because the contribution of background to the magnitude of the peak (which, unfortunately, can not be obtained from the Fig. 3) must be taken into account.

This outcome is quite natural and to be expected. Alkali and alkaline earth cations are not the only mobile constituents in glass responding to the mechanical perturbation. The main component of all oxide glasses is oxygen. It comprises about 70 wt.% of glass composition and more than 90% of its specific volume, and it can be found in glass in various structural forms: bridging and non-bridging, peroxide linkages and radicals, hydroxyl groups and oxygen molecules and atoms. Activation energies characterizing the oxygen transport in glass are very close to those retrieved from the mechanical loss measurements (see for example [41,42]). These data have been available for decades but, unfortunately, have not received the attention which they merit.

CONCLUSION

This work continues the exploration of the defect model for the mixed mobile ion effect. It provides consistent and satisfactory explanation for the peculiar features of the MCE in mixed alkaline earth and alkali alkaline earth glasses. The quantitative analysis of the mechanical loss spectra of single alkali, mixed alkali and alkali alkaline earth glasses performed here yields a much clearer picture of the sub-$T_g$ mechanical relaxation processes in glass and allows to explicitly link the mechanical loss maxima observed between ~ 200$^O$C and $T_g$ to the movement of oxygen-related structural moieties.